\documentclass[aps,amssymb,amsfonts,twocolumn,showpacs]{revtex4}
\usepackage{graphicx} \usepackage{bm}
 \usepackage{amsmath}
 \usepackage{amssymb}
 \usepackage{latexsym}
 \usepackage{amsfonts}
 \usepackage{epsfig}

\newcommand\br{\mathbf{r}}

\newcommand\bv{\mathbf{v}}

 \newcommand{\bea}{\begin{eqnarray}}
 \newcommand{\ea}{\end{eqnarray}}
 \newcommand{\eea}{\end{eqnarray}}
 
\newcommand{\edd}{\varepsilon_{\rm dd}}
 
\newcommand{\ltsimeq}{\raisebox{-0.6ex}{$\,\stackrel
        {\raisebox{-.2ex}{$\textstyle <$}}{\sim}\,$}}
\newcommand{\gtsimeq}{\raisebox{-0.6ex}{$\,\stackrel
        {\raisebox{-.2ex}{$\textstyle >$}}{\sim}\,$}}

\begin{document}
\title{Structure formation during the collapse of a dipolar atomic Bose-Einstein condensate}
\author{N.G. Parker$^{1}$, C. Ticknor$^2$, A.M. Martin$^3$ and D.H.J. O'Dell$^1$}
\affiliation{$^1$ School of Physics and Astronomy, McMaster
University, Hamilton, Ontario, L8S 4M1, Canada
\\ $^2$ ARC Centre of Excellence for Quantum-Atom Optics and Centre for Atom Optics and Ultrafast Spectroscopy, Swinburne University of Technology, Hawthorn, Victoria 3122, Australia
\\ $^3$ School of Physics, University of Melbourne, Parkville, VIC 3010, Australia}

\begin{abstract}
We investigate the collapse of a trapped dipolar Bose-Einstein
condensate.  This is performed by numerical simulations of the
Gross-Pitaevskii equation and the novel application of the
Thomas-Fermi hydrodynamic equations to collapse.  We observe regimes
of both global collapse, where the system evolves to a highly
elongated or flattened state depending on the sign of the dipolar
interaction, and local collapse, which arises due to dynamically
unstable phonon modes and leads to a periodic arrangement of density
shells, disks or stripes.  In the adiabatic regime, where ground
states are followed, collapse can occur globally or locally, while
in the non-adiabatic regime, where collapse is initiated suddenly,
local collapse commonly occurs.  We analyse the dependence on the
dipolar interactions and trap geometry, the length and time scales
for collapse, and relate our findings to recent experiments.

\end{abstract}


\maketitle Wavepacket collapse is a phenomenon seen in diverse
physical systems whose common feature is that they obey non-linear
wave equations \cite{collapse}, e.g., in nonlinear optics
\cite{kivshar}, plasmas \cite{plasmas} and trapped atomic
Bose-Einstein condensates (BECs)
\cite{bradley,gerton00,roberts01,chin03,koch,lahaye}. In the latter
case, collapse occurs when the atomic interactions are sufficiently
attractive. For the usual case of isotropic $s$-wave interactions
experiments have demonstrated both global \cite{gerton00} and local
collapse \cite{chin03} depending upon, respectively, whether the
imaginary healing length is of similar size or much smaller than the
BEC \cite{sackett98}. During global collapse the monopole mode
becomes dynamically unstable and the BEC evolves towards a point
singularity, with the threshold for collapse generally exhibiting a
weak dependence on trap geometry \cite{gammal,parker}.  Local
collapse occurs when a phonon mode is dynamically unstable such that
the collapse length scale is considerably smaller than the BEC.

Recently, the Stuttgart group demonstrated collapse in a BEC with
\emph{dipole-dipole} interactions, where the atomic dipoles were
polarized in a common direction by an external field
\cite{koch,lahaye}. The long-range nature of dipolar interactions
means that the Gross-Pitaevskii wave equation that governs the BEC
is not only non-linear but also non-local
\cite{goral,yi,santos00,yi01}.  On top of being long-range, dipolar
interactions are also anisotropic, being attractive in certain
directions and repulsive in others.  This anisotropy has manifested
itself experimentally in the stability of the ground state, which is
strongly dependent on the trap geometry \cite{koch}, and in the
anisotropic collapse of the condensate \cite{lahaye}.  Some
uncertainty exists over the mechanism of collapse in these systems.
In the latter experiment, striking images indicate that the
condensate underwent global collapse, which is likely to have
occurred through a {\em quadrupole} mode
\cite{yi01,goral02,ticknor}.  In the former experiment, however,
recent theoretical results suggest that local collapse played a
dominant role \cite{bohn}.

A unique feature of trapped dipolar BECs in comparison to {\it
s}-wave BECs is that they are predicted to exhibit minima in their
excitation spectrum at finite values of the excitation quantum
number
\cite{odell03,santos03,giovanazzi04a,fischer06,ronen07,ronen08},
reminiscent of the roton minimum in the quantum liquid He-II.  For
gaseous BECs the depth of the minimum is tunable via microscopic
parameters such as the dipolar and {\it s}-wave interaction
strengths. An important physical consequence of the ``roton''
minimum is that it can lead to density modulations in the ground
state dipolar BEC
\cite{giovanazzi02a,odell03,giovanazzi04a,ronen07,dutta07,dutta07b,ronen08}.
However, the regions of parameter space where they occur are small
and lie close to the unstable region \cite{ronen07,dutta07b}. It is
therefore natural to ask if these characteristic density modulations
form during collapse where they might be more readily visible. As we
shall see, collapse experiments can indeed provide us with an
indirect yet accessible way of studying such effects.

The dipolar BECs that have been realized thus far
\cite{griesmaier,gorceix} are formed of $^{52}$Cr atoms with
magnetic dipole moments $d$ and coupling strength $C_{\rm dd}=\mu_0
d^2$, where $\mu_0$ is the permeability of free space.  Ultracold
quantum gases of polar molecules, which feature electric dipolar
interactions, are also likely to be formed in the near future
\cite{jin}. The dipolar interactions typically co-exist with {\it
s}-wave interactions of characteristic amplitude
$g=4\pi\hbar^2a_{\rm s}/m$, where $a_{\rm s}$ is the {\it s}-wave
scattering length, and an important parameter is the ratio
\cite{varepsilon},
\begin{equation}
 \varepsilon_{\rm dd}=C_{\rm dd}/3g.
 \label{eq:edddefinition}
\end{equation}
The {\it s}-wave coupling $g$ can be effectively tuned between
positive and negative infinity using a Feshbach resonance, and this
was employed to control the stability of the dipolar BEC in the
recent experiments \cite{koch} and \cite{lahaye}.  Both the
amplitude and sign of the dipolar coupling $C_{\rm dd}$ can also be
tuned  by rotation of the polarization axis \cite{giovanazzi02}. A
huge parameter space of dipolar interactions, from $\varepsilon_{\rm
dd}=-\infty$ to $+\infty$, is thereby accessible for study.

Insight into the collapse instability of a trapped dipolar BEC can
be obtained by considering its ground states.  Such a route could be
followed by beginning with a stable ground state and adiabatically
tuning the parameters towards the instability, which we term {\em
adiabatic collapse}.  Consider tuning the ratio $\edd$: as it is
increased the ground state density profile evolves so as to benefit
from the attractive part of the interaction.  This can happen both
globally and locally.  For example, when $C_{\rm dd}>0$ the system
can undergo global magnetostriction and elongate along the axis of
polarization \cite{goral,santos00,yi01,odell04,eberlein05}, tending
towards a collapsed state of a line of end-to-end dipoles. However,
the dipoles can also rearrange themselves locally
\cite{giovanazzi02a,odell03,giovanazzi04a,ronen07,dutta07,dutta07b,ronen08}
and in particular, in a pancake-shaped geometry for $C_{\rm dd}>0$ a
``red blood cell'' density profile has been predicted
\cite{ronen07}.

If collapse is triggered suddenly, which we term {\em non-adiabatic
collapse}, the system does not follow the ground state solutions and
excitations play a role.  In the dipolar collapse experiment
\cite{lahaye} the collapse was initiated by a change in $g$ which
took place over approximately one trap period.  This time scale lies
on the border between adiabatic and non-adiabatic collapse.

Motivated by these issues we study theoretically the global and
local collapse of a dipolar BEC over a significant range of $\edd$
that is accessible to current experiments.  Our analysis is based on
simulations of the Gross-Pitaevskii equation and the hydrodynamic
(Thomas-Fermi) approximation, including the novel application of the
hydrodynamic equations of motion to collapse.  We observe that
collapse occurs anisotropically with the dipoles tending to align
themselves end-to-end for $C_{\rm dd}>0$ and side-by-side for
$C_{\rm dd}<0$. If collapse is approached adiabatically it occurs
globally for moderate dipolar interactions ($-1 \ltsimeq \edd
\ltsimeq 2$) and beyond this we have also observed signs of local
collapse.  When collapse is initiated suddenly, it is dominated by
the formation of local density structures, whose shape is determined
by the dipolar interactions and can be related to unstable
Bogoliubov modes.  We map out the length and time scales for
collapse, and the role of interaction strength and trap geometry. We
then compare our results to recent experiments, in particular that
of Lahaye {\it et al}.\ \cite{lahaye}, where the condensate appeared
to undergo global collapse.  We show that our results are consistent
with this observation and, furthermore, indicate how local collapse
could be induced in this current experimental set-up.

In Section \ref{sec:theory} we introduce the mean-field description
of dipolar BECs and, by employing the Thomas-Fermi approximation,
derive the static solutions of the system and hydrodynamic equations
of motion.  In Section \ref{sec:solutions} we discuss the static
solutions and their threshold for collapse. In Section
\ref{sec:collapse} we consider non-adiabatic collapse, induced by a
sudden change in the interaction strength, and compare the
hydrodynamic predictions with simulations of the Gross-Pitaevskii
equation.  In Section \ref{sec:length_time}, we extend our analysis
of non-adiabatic collapse to map out the time and length scales of
collapse, and in the latter case, show that the homogeneous
Bogoliubov spectrum gives good agreement with the observed local
collapse.  In Section \ref{sec:disc} we relate our findings to
recent collapse experiments \cite{koch,lahaye} and in Section
\ref{sec:concs} we present our conclusions.

\section{Theoretical framework}
\label{sec:theory}
\subsection{Dipolar Gross-Pitaevskii equation}
For weak interactions and at zero temperature the mean-field ``wave function'' of an atomic BEC $\psi \equiv \psi({\bf r},t)$ satisfies the Gross-Pitaevskii equation
(GPE),
\begin{equation}
i\hbar \frac{\partial \psi}{\partial
t}=\left(-\frac{\hbar^2}{2m}\nabla^2+V({\bf r})+g|\psi|^2+\Phi_{\rm
dd}({\bf r},t)\right)\psi. \label{eqn:GPE}
\end{equation}
We assume that the confining potential has the cylindrically-symmetric harmonic form,
\begin{equation}
 V_{\rm ext}({\bf r})=\frac{1}{2}m\omega_x^2(x^2+y^2+\gamma^2 z^2),
\label{eq:trap}
\end{equation}
where $\omega_x$ is the radial trap frequency and
$\gamma=\omega_z/\omega_x$ is the trap's aspect ratio (which will be
henceforth termed the {\em trap ratio}). Note that the trap has
axial and radial harmonic oscillator lengths
$a_z=\sqrt{\hbar/m\omega_z}$ and $a_x=\sqrt{\hbar/m\omega_x}$.
Dipolar atomic interactions are described by the non-local
mean-field potential $\Phi_{\rm dd}$ \cite{yi,goral,santos00,ronen},
\begin{equation}
\Phi_{\rm dd}({\bf r})=\int d^3{\bf r}' U_{dd}({\bf r}-{\bf
r}')|\psi({\bf r}')|^2.
\label{eqn:phidd1}
\end{equation}
The interaction potential between two dipoles separated by $\br$ and
aligned by an external field along a unit vector $\hat{\mathbf{e}}$
is given by,
\begin{equation}
U_{\mathrm{dd}}(\br)= \frac{C_{\mathrm{dd}}}{4 \pi}\, \hat{{\rm
e}}_{i} \hat{{\rm e}}_{j} \frac{\left(\delta_{i j}- 3 \hat{r}_{i}
\hat{r}_{j}\right)}{r^{3}} . \label{eqn:Udd}
\end{equation}
Throughout this work we consider the dipoles to be aligned in the
{\it z}-direction. It is useful to specify the dipolar interaction
strength in Eq.~(\ref{eqn:GPE}) by the parameter,
\begin{equation}
 k_{\rm dd}=Na_{\rm dd}/a_x,
\end{equation}
where $a_{\rm dd}=C_{\rm dd}m/(12\pi\hbar^2)$ is the dipolar
``scattering length'' and $N$ is the total number of condensate
atoms. Note that the ratio $\edd$ can be written as
$\edd=a_{\mathrm{dd}}/a_{\rm s}$. In the dipolar BEC collapse
experiments  \cite{koch,lahaye} $k_{\rm dd}$ lies in the range
$25-50$.  We will assume throughout this work that $g>0$ such that
the case of $\edd<0$ corresponds to $C_{\rm dd}<0$. While the
opposing case of $g<0$ is experimentally accessible, the {\it
s}-wave interactions will typically induce collapse, rather than the
dipolar interactions, and so will not be considered in this work.

It is important to note that the basic GPE is insufficient to
describe the full collapse dynamics since higher-order effects, e.g.
three-body loss, can become significant as the density escalates.
However, the GPE provides an excellent prediction for the {\em
onset} of collapse \cite{parker,gammal,koch} and can be expected to
accurately describe the early collapse dynamics. Our study will
therefore consider the dynamics up to this point. Extension of these
results to the full collapse dynamics could be made in future by
including a three-body loss term in the GPE \cite{lahaye,ticknor}.

\subsection{Thomas-Fermi solutions}
Static solutions of the GPE can be expressed as $\psi({\bf
r},t)=\psi_0({\bf r})\exp[-i\mu t/\hbar]$, where $\mu$ is the
chemical potential.  We enter the Thomas-Fermi (TF) regime when the
interaction energy dominates over the energy arising from density
gradients, termed the zero-point kinetic energy \cite{dalfovo}.  We
then neglect the zero-point energy and the atomic density
$n_0=|\psi_0|^2$ satisfies the expression,
\begin{equation}
V({\bf r})+g n_0({\bf r})+\Phi_{dd}({\bf r})=\mu \label{eqn:tiGPE}
\end{equation}

For an {\it s}-wave BEC, the ratio of interaction energy to
zero-point energy is commonly specified as $k_{\mathrm{s}}=Na_{\rm
s}/a_x$, with the system entering the TF regime when $k_{\mathrm{s}}
\gg 1$. The criterion for the TF regime is not so simple for a
dipolar BEC since the anisotropic interactions make it strongly
dependent on its shape. However, in the limiting cases of a highly
elongated ``cigar'' condensate, or a highly flattened ``pancake''
condensate, the TF regime is valid when, respectively
\cite{Parker08},
\begin{eqnarray}
 \frac{\gamma k_{\rm dd}}{\edd}(1-\edd) & \gg & 1 \quad \mbox{[cigar]} \label{eq:cigarTF}\\
 \frac{k_{\rm dd}}{\gamma^{3/2} \edd}(1+2\edd) & \gg & 1 \quad \mbox{[pancake]} \label{eq:pancakeTF}.
\end{eqnarray}

A solution of Eq.~(\ref{eqn:tiGPE})  is given by an inverted
parabola of the form \cite{odell04},
\begin{eqnarray}
n_0(\br) &=& \frac{15 N}{8\pi R_{x0} R_{y0} R_{z0}}
\left[1-\frac{x^2}{R_{x0}^2}-\frac{y^2}{R_{y0}^2}
 - \frac{z^2}{R_{z0}^2}\right], \label{parabola1}
\end{eqnarray}
valid where $n_0(\br) \geq0$ and $n_0(\br)=0$ elsewhere, and the TF
radii of the condensate are denoted by $R_{x0}$, $R_{y0}$ and
$R_{z0}$. Note that the $0$-subscript denotes static solutions.
When the trap is cylindrically-symmetric about the same direction as the polarizing field, as we assume here, the density is also cylindrically symmetric.  Its aspect ratio $\kappa_0=R_{x0}/R_{z0}$ satisfies a
transcendental equation \cite{yi01,odell04},
\begin{equation}
 \frac{\kappa_{0}^{2}}{\gamma^{2}}
\left[\frac{3 \varepsilon_{\mathrm{dd}}
f(\kappa_{0})}{1-\kappa_{0}^{2}} \left(
\frac{\gamma^{2}}{2}+1\right)-2\varepsilon_{\mathrm{dd}} -1 \right]
 =
\varepsilon_{\mathrm{dd}}-1. \label{eq:transcendental}
\end{equation}
where,
\begin{equation}
f(\kappa)=\frac{1+2\kappa^2}{1-\kappa^2}-\frac{3\kappa^2
\mathrm{arctanh}\sqrt{1-\kappa^2}}{\left(1-\kappa^2\right)^{3/2}}
\quad \label{eq:f}.
\end{equation}
The equilibrium radii are given by \cite{odell04},
\begin{eqnarray}
R_{x0}=R_{y0}=\left[\frac{15 N g \kappa_{0}}{4\pi m \omega_x^2}
\left\{1+ \varepsilon_{\mathrm{dd}} \left( \frac{3}{2}
\frac{\kappa_{0}^{2} f(\kappa_{0})}{1-\kappa_{0}^{2}}-1 \right)
 \right\} \right]^{1/5} \label{eq:Rxsol}
\end{eqnarray}
and $R_{z0}=R_{x0}/\kappa_{0}$.  For an {\it s}-wave BEC, we retrieve the simple and expected result that $\kappa_0=\gamma$.  The anisotropic interactions, however, lead to magnetostriction such that $\kappa_0<\gamma$ for $\edd>0$ and $\kappa_0>\gamma$ for $\edd<0$.

Consider the broader range of cylindrically-symmetric density
profiles which have the same form as Eq.~(\ref{parabola1}) but are
not limited to the equilibrium solutions. These general profiles
define an energy ``landscape'' in terms of $R_x$ and $\kappa$ given
by \cite{eberlein05},
\begin{eqnarray}
\frac{E(R_x,\kappa)}{N}=\frac{m \omega_{x}^{2}R_{x}^{2}}{14}
\left(2+\frac{\gamma^{2}}{\kappa^{2}}\right)+\frac{15Ng\kappa}{28\pi
R_{x}^{3}}\left[1-\varepsilon_{\mathrm{dd}} f(\kappa)\right].
\label{eq:E}
\end{eqnarray}
Then the static solutions (\ref{eq:transcendental}) correspond to
stationary points in this landscape located at $(R_{x0},\kappa_0$).
By examining the landscape in the vicinity of the solution one can
determine whether it is stable (a global or local energy minimum) or
unstable (maximum or saddle point).

The TF regime can be formally regarded as the $N \rightarrow \infty$
limit of Gross-Pitaevskii theory. Indeed, the TF predictions for the
stability of a dipolar BEC depend only upon $\edd$ and $\gamma$,
thus simplifying the parameter space.  However the TF model does not
accurately describe situations where the zero-point energy is
considerable, e.g. close to the collapse threshold. In order to
discern the effect of the zero-point energy we shall compare the TF
solutions with those of a variational approach based upon a gaussian
ansatz, as detailed in Appendix A, which includes this energy
contribution \cite{santos00,yi01}.    To describe deviations from
the TF (or gaussian) solutions, e.g.``red-blood cell'' states
\cite{ronen07}, one must solve the full GPE.  This will also be
considered in due course.

\subsection{Thomas-Fermi equations of motion for collapse}
To enable a hydrodynamic interpretation we employ the Madelung
transform $\psi({\bf r},t)=\sqrt{n({\bf r},t)}\exp[iS({\bf r},t)]$,
where $n({\bf r},t)$ and $S({\bf r},t)$ are the density and phase
distributions, and define the ``fluid'' velocity as $v({\bf
r},t)=(\hbar/m)\nabla S({\bf r},t)$ \cite{madelung}. In the TF limit
the dipolar GPE then leads to hydrodynamic equations \cite{dalfovo},
\begin{eqnarray}
\frac{\partial n}{\partial t} &=& - {\bf \nabla} \cdot (n \bv) \;,\label{continuity}\\
m\frac{\partial {\bf v}}{\partial t} &=&- {\bf \nabla} \left( \frac
{ m v^2}{2} + V + g n + \Phi_{\mathrm{dd}}\right)
\;.\label{euler}
\end{eqnarray}
Following earlier work for {\it s}-wave BECs \cite{kagan,castin}
there exists a class of exact time-dependent scaling solutions to
Eqs.~(\ref{continuity}) and (\ref{euler}) given by \cite{odell04},
\begin{eqnarray}
n(\br,t) &=& n_0(t)
\left[1-\frac{x^2}{R_x^2(t)}-\frac{y^2}{R_y^2(t)}
 - \frac{z^2}{R_z^2(t)}\right] \label{parabola2} \\
\bv(\br,t) & = & \frac{1}{2} \mathbf{\nabla} \left[ \alpha_x(t)
x^{2} + \alpha_y(t) y^{2}+ \alpha_z(t) z^{2} \right]
\label{velocity}
\end{eqnarray}
valid where $n(\br,t) \geq0$ and $n(\br,t)=0$ elsewhere, and
$n_0(t)=15 N/(8\pi R_x(t) R_y(t) R_z(t))$ is the peak density. The
time evolution of the radii $R_j$ is governed by three
\emph{ordinary} differential equations, with the components of the
velocity field given by $\alpha_{j}= \dot{R}_{j}/R_{j}$, where
$j=x,y,z$.

Restricting ourselves to cylindrically-symmetric dynamics
where $R_{x}(t)=R_{y}(t)$, and introducing the scaling factors
$\lambda_{x}(t)=R_x(t)/R_{x0}$ and $\lambda_{z}(t)=R_z(t)/R_{z0}$
(recall that the $0$-subscript denotes the initial static solution), the
time evolution is determined by the coupled ordinary differential equations,
\begin{eqnarray}
\ddot{\lambda}_{x} & = & - \omega_x^2 \lambda_{x} + \; \frac{\eta
g(t) \kappa_{0}}{\lambda_{x}
\lambda_{z}}\bigg[\frac{1}{\lambda_{x}^2} \nonumber \\
 & & \mbox{ \hspace{1em}}
-\varepsilon_{\mathrm{dd}}(t)
\left(\frac{1}{\lambda_{x}^2}+\frac{3}{2} \kappa_{0}^{2}\frac{
f(\kappa_{0}\lambda_{x}/\lambda_{z})}{\kappa_{0}^{2}\lambda_{x}^{2}-\lambda_{z}^{2}
}\right) \bigg]
\label{eq:rxmotion} \\
\ddot{\lambda}_{z}& = & -\omega_x^2 \gamma^{2}\lambda_{z}+
\;\frac{\eta g(t)\kappa_{0}^{3}}{\lambda_{x}^{2}}
\bigg[\frac{1}{\lambda_{z}^{2}} \nonumber \\
 & & \mbox{ \hspace{1em}} +2\varepsilon_{\mathrm{dd}}(t) \left(\frac{1}{\lambda_{z}^2}+\frac{3}{2}
\frac{f(\kappa_{0} \lambda_{x}/\lambda_{z})}{\kappa_{0}^{2}
\lambda_{x}^{2}-\lambda_{z}^{2} }\right) \bigg]. \label{eq:rzmotion}
\end{eqnarray}
where $\eta=15N/4\pi m R_{x0}^5$.  These equations are significantly
less demanding to solve than the full GPE and in certain limits
analytic solutions exist (see Eq.\ (\ref{eq:lamxsol}) below). The TF
equations of motion have been successfully applied to model the
condensate dynamics under time-dependent trapping including the
important case of ballistic expansion \cite{dalfovo,giovanazzi06}.
They describe two independent collective excitation modes of the
system: the monopole mode (when $\lambda_{x}(t)$ and
$\lambda_{z}(t)$ are in phase) and the axis-symmetric quadrupole
mode (when $\lambda_{x}(t)$ and $\lambda_{z}(t)$ are $180^{\circ}$
out of phase). In this paper we employ the TF equations of motion to
study global collapse.  In the pure {\it s}-wave case global
collapse occurs through the monopole mode \cite{sackett98}, but when
dipolar interactions dominate global collapse occurs through the
quadrupole mode \cite{yi01,goral02,ticknor}.

We will employ the TF equations of motion to describe non-adiabatic
collapse, triggered by a sudden change in $\varepsilon_{\rm dd}(t)$.
To conform to the current experimental method \cite{koch,lahaye} we
shall implement this through a sudden change in the {\it s}-wave
interactions $g(t)$. By, (i) starting with the BEC well below the
threshold for collapse and (ii) suddenly changing to a state which
is well above the threshold for collapse, the regime where
zero-point energy dominates can be bypassed and the TF equations
should apply throughout.  We will verify this in due course.

\section{Adiabatic collapse}
\begin{figure}[b]
\centering
\includegraphics[width=7cm,clip=true]{Fig1a.eps}
\includegraphics[width=8.5cm,clip=true]{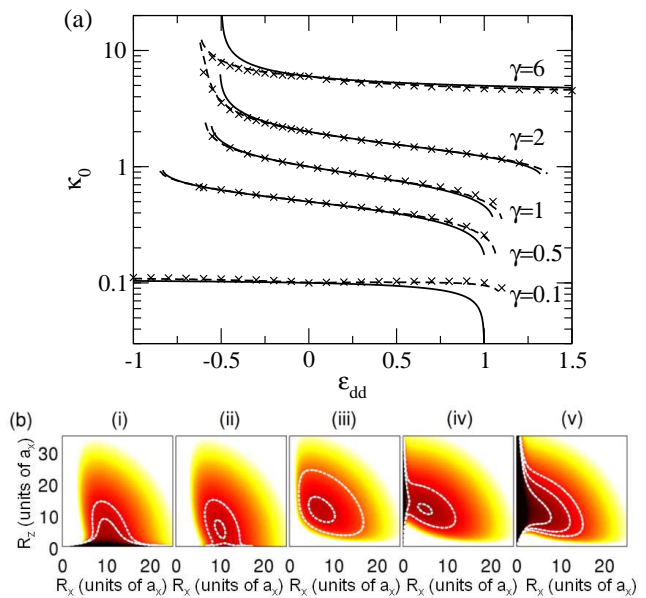}
\caption{(Color online)(a) Aspect ratio $\kappa_0$ of the ground
state solutions as a function of $\varepsilon_{\rm dd}$ for various
trap ratios $\gamma$. Presented are the stable TF solutions of
Eq.~(\ref{eq:transcendental}) (solid lines), GPE solutions with
$k_{\rm dd}=115$ (crosses), and variational solutions for $k_{\rm
dd}=115$ (dashed line). (b) Energy landscapes of Eq.~(\ref{eq:E})
for $\gamma=2$ and $\varepsilon_{\rm dd}=$ (i) $-0.8$, (ii) $-0.52$,
(iii) $0.8$, (iv) $1.025$ and (v) $1.4$. Light/dark regions
correspond to high/low energy, and white contours help to visualise
the landscapes.} \label{fig:kappa}
\end{figure}
\label{sec:solutions}

Imagine an experiment that starts with a stable ground state BEC and
adiabatically increasing the magnitude of $\edd$. The condensate
will remain in the ground state corresponding to the instantaneous
value of $\edd$ and will eventually collapse at a critical value of
$\edd$ \cite{goral,santos00}. The threshold for collapse in general
depends upon $\edd$, $g$, $\gamma$ and $N$.  We consider two
possible scenarios for adiabatic collapse: i) if there is no roton
minimum in the excitation spectrum then adiabatic collapse proceeds
in a similar manner to the usual pure {\it s}-wave case, i.e.\ a
{\em global collapse} via a low-lying shape oscillation mode once
the (imaginary) healing length becomes of the same order as the
condensate size \cite{sackett98}; ii) if there is a roton minimum
then this deepens as $\edd$ increases and, at the point at which it
reaches zero energy, can lead to {\em local collapse} on a length
scale determined by the roton minimum. We note that technically
speaking neither scenario can be truly adiabatic since the mode
responsible for collapse has zero frequency at the collapse
threshold, but the non-adiabaticity can be confined to a small
region of parameter space if $\edd$ is increased slowly enough.

\subsection{Global adiabatic collapse}
Ground state solutions, characterised by their aspect ratio
$\kappa_0$, are presented in Fig.~\ref{fig:kappa}(a) as a function
of $\varepsilon_{\rm dd}$ for various trap ratios $\gamma$. The
figure shows the predictions given by numerical solution of the GPE,
the parabolic TF solution, and the gaussian variational ansatz. We
first consider the parabolic TF solutions (solid lines in
Fig.~\ref{fig:kappa}(a)), which, we recall, can be characterized
solely by $\edd$ and $\gamma$. For $\varepsilon_{\rm dd}=0$ we
observe that $\kappa_0=\gamma$, as expected for {\it s}-wave
condensates. We will consider the regimes of positive and negative
$\edd$ separately with the aid of typical energy landscapes pictured
in Fig.~\ref{fig:kappa}(b):

\begin{itemize}
 \item{{\bf $\varepsilon_{\rm dd}>0$}: As $\edd$ is increased from zero, $\kappa_0$ decreases since the dipoles prefer to lie end-to-end along $z$.  The experiment \cite{lahaye_ferro} observed this magnetostrictive effect, in good agreement with the TF
predictions. For $0<\edd<1$, the parabolic solutions
(\ref{parabola1}) are global minima of the TF energy functional
\cite{eberlein05} (Fig.~\ref{fig:kappa}(b)(iii)).  For $\edd>1$,
however, the TF solution becomes only a local minimum
(Fig.~\ref{fig:kappa}(b)(iv)), with the global minimum being a
collapsed state of zero width.  Indeed, there is an upper critical
dipolar-to-{\it s}-wave ratio, $\edd^{c+}\geq1$, beyond which the
local minimum disappears and the whole system is unstable to
collapse into a $\kappa_0=0$ state (Fig.~\ref{fig:kappa}(b)(v)),
i.e. an infinitely thin line of dipoles.  Note that  $\edd^{c+}$
depends on the trap ratio $\gamma$.  Elongated BECs, being
predominantly attractive, are most unstable, with $\varepsilon_{\rm
dd}^{c+}\approx1$. Increasing $\gamma$ increases $\varepsilon_{\rm
dd}^{c+}$ due to the increasing repulsive interactions in the
system. Indeed, for $\gamma\gtsimeq5.17$, $\varepsilon_{\rm
dd}^{c+}=\infty$ \cite{santos00,yi01,eberlein05}, i.e., the
parabolic solutions are robust to collapse for any interaction
strength (see the case of $\gamma=6$ in Fig.~\ref{fig:kappa}(a)).}
\item{{\bf $\varepsilon_{\rm dd}<0$}: As $\edd$ is decreased from zero,
$\kappa_0$ increases since the dipoles now prefer to lie
side-by-side in the transverse plane.  For $-0.5<\edd<0$, the
parabolic solutions are robust to collapse, while for $\edd<-0.5$,
they become local energy minima with the global minima being a
collapsed plane of dipoles (Fig.~\ref{fig:kappa}(b)(ii)).  We define
a critical value $\varepsilon_{\rm dd}^{c-}\leq-0.5$, below which no
stable parabolic solutions exist (Fig.~\ref{fig:kappa}(b)(i)) and
the system is unstable to collapse into a $\kappa_0=\infty$ state,
i.e., an infinitely thin plane of side-by-side dipoles.  Pancake
geometries are particularly prone to this with $\varepsilon_{\rm
dd}^{c-}\approx -0.5$, while the system becomes increasingly stable
in more elongated geometries.  Indeed, in sufficiently elongated
geometries $\gamma\ltsimeq0.19$, the parabolic solutions are stable
to collapse with $\varepsilon_{\rm dd}^{c-}=-\infty$.}
\end{itemize}

Numerical solutions of the time-independent GPE for $k_{\rm dd}=115$
are shown in Fig.~\ref{fig:kappa}(a) as crosses.  Our method of
determining the BEC widths is detailed in Appendix B. While the GPE
solutions are generally in good agreement with the TF results,
deviations become significant near the point of collapse where the
zero-point energy has a considerable stabilising effect on the
solutions. We have additionally performed time-dependent simulations
in which $\varepsilon_{\rm dd}$ is varied slowly, and find that the
condensate follows these ground state solutions and maintains a
parabolic-like density profile.  This is consistent with global
collapse.  We have performed a similar analysis for $k_{\rm dd}=14$
and observe the same qualitative results, with the threshold for
collapse pushed to slightly higher values of $|\edd|$ due to
enhanced zero-point energy.

The dashed lines in Fig.~\ref{fig:kappa}(a) are the results given by
the gaussian variational method of Appendix A.  Note that if
zero-point effects were neglected, the gaussian solutions would
satisfy the same transcendental equation (\ref{eq:transcendental})
as the parabola, i.e., the aspect ratio of the BEC $\kappa$ is
largely independent of the ansatz (although the profile and radii do
differ).  Indeed, for small values of $|\edd|$ the variational and
TF predictions are almost identical.  However, close to the onset of
collapse these predictions deviate.  Importantly, the gaussian
variational method gives excellent agreement with the full GPE
solutions right up to the point of collapse, highlighting the
importance of zero-point energy at the point of collapse.

\subsection{Local adiabatic collapse}
A surprising result of the parabolic density profiles is that within
the TF approximation they remain stable even as $\edd \rightarrow
\infty$ and $\edd \rightarrow -\infty$, providing the trap ratio is
sufficiently extreme ($\gamma>5.17$ for $\edd>0$ and $\gamma<0.19$
for $\edd<0$). Similar behaviour arises for a variational approach
based on a gaussian density profile \cite{santos00,yi01}. The recent
experiment of Koch {\it et al.} \cite{koch} has observed the
stability of a dipolar condensate under various trap ratios, in
reasonable qualitative agreement with the gaussian and TF
predictions. However, numerical solutions of the GPE show that, even
though the stability is enhanced under extreme trap ratios, collapse
will occur for strong enough dipolar interactions \cite{yi01}. This
apparent contradiction arises because the gaussian and parabolic
solutions are only capable of describing low-lying monopole and
quadrupole fluctuations. In the presence of a roton minimum the
instability of a mode with high quantum number can lead to local
adiabatic collapse.  Bohn {\it et al.} recently employed a
theoretical model that allowed for local collapse and found improved
quantitative agreement with the experimental observations in pancake
geometries \cite{bohn}, suggesting that local collapse does indeed
play a key role in such geometries. Note that, by contrast, in {\it
s}-wave condensates an adiabatic reduction in $g$ should always
induce global collapse.

In order to induce local adiabatic collapse it is necessary to use
values of $\edd$ that fall outside $-1 \ltsimeq\edd\ltsimeq2$ (the
region of Fig.~1). We have probed pancake-shaped ground state
solutions that extend beyond this range.  For $\gamma=10$ and $20$
we have probed up to $\edd=30$ with no evidence for collapse.
However, for $\gamma=6$ we have observed the adiabatic onset of a
local collapse instability at $\edd \approx 2.2$, characterised by
the formation and growth of cylindrical density shells (similar to
those that will be discussed in Section \ref{sec:collapse}). Our
results are consistent with those of Ronen {\it et al.}
\cite{ronen07}, who predicted that, in a purely dipolar BEC
($\edd=\infty$), collapse is possible for large $\gamma$ and that,
close to the collapse threshold, the ground state adopts density
corrugations. It is reasonable to assume that as one passes into the
unstable region that these lead to local collapse.  This picture was
also recently suggested by Bohn {\it et al.} \cite{bohn}. Note that
for the moderate range of $\edd$ that we concentrate on here, the TF
solutions are a very good approximation to the Gross-Pitaevskii
solutions and the adiabatic collapse proceeds globally.

\section{Non-adiabatic collapse}
\begin{figure}[b]
\centering
\includegraphics[width=8.5cm,clip=true]{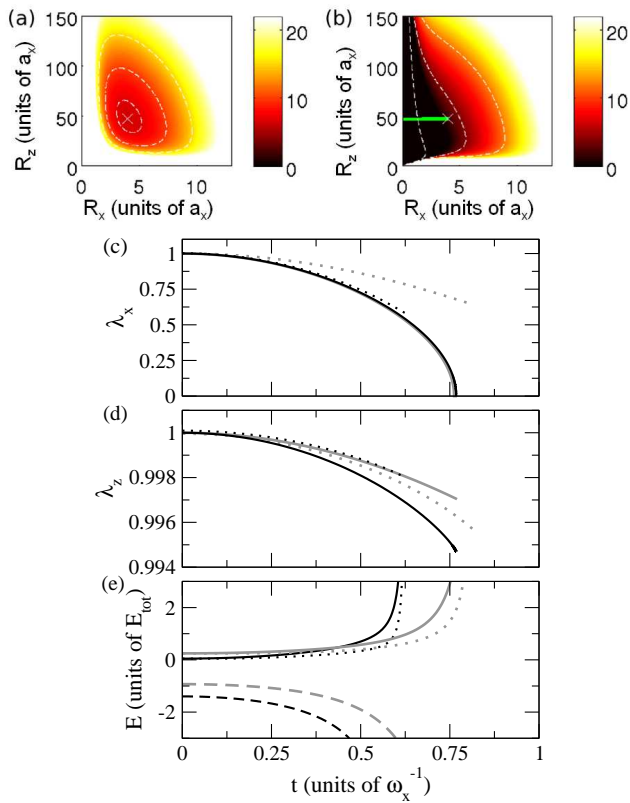}\\
\includegraphics[width=6cm,clip=true]{Fig2b.eps}\\
\caption{(Color online) Collapse dynamics in a cigar trap
$\gamma=0.1$ for $\varepsilon^0_{\rm dd}=0.8$ and $\varepsilon^{\rm
f}_{\rm dd}=1.4$. (a) Energy landscape of Eq.~(\ref{eq:E}) at $t=0$.
(b) Energy landscape for $t>0$, with the ensuing TF trajectory
indicated (green/grey line).  Energy is scaled in units of
$\hbar\omega_x$. (c) $\lambda_x(t)$ and (d) $\lambda_z(t)$ from the
full TF equations of motion (black solid line), the simplified case
of Eqs.~(\ref{eq:lamxeqn},\ref{eq:lamzeqn}) (solid grey line), and
GPE simulations for $k_{\rm dd}=80$ (grey dotted line) and $2000$
(black dotted line). (e) Evolution of the energy components during
GPE simulations with $k_{\rm dd}=80$ (grey lines) and $k_{\rm
dd}=2000$ (black lines).  Shown are the kinetic $E_{\rm k}$ (solid
line), zero-point $E_{\rm zp}$ (dotted line) and dipolar $E_{\rm d}$
(dashed line) energies, renormalised by the total energy $E_{\rm
tot}$.  } \label{cigar_evo1}
\end{figure}
\label{sec:collapse}

An alternative way to induce collapse is to perform a non-adiabatic
change of $\edd$ of the form,
\begin{equation}
 \edd=\left\{
\begin{array}{c}
\varepsilon^0_{\rm dd}\\
\varepsilon^{\rm f}_{\rm dd}
\end{array}
\right.
{\rm for}
\begin{array}{cr}
 & t=0 \\
 & t>0
\end{array}\label{eq:time}
\end{equation}
We will assume that this is achieved by tuning the {\it s}-wave interactions from $g_{0}=C_{\rm dd}/3\varepsilon^0_{\rm dd}$ to
$g_f=C_{\rm dd}/3\varepsilon^{\rm f}_{\rm dd}$.

In order for the TF equations of motion to remain valid we require
that the TF approximation holds throughout the dynamics i.e.\
regimes where the zero-point energy is important are avoided. This
requires that the initial condensate is well inside the regime of
stable solutions ($|\edd^0|<|\edd^c|$), and
subsequently that the system is deep within the collapse regime
($|\edd^{\rm f}|>|\edd^c|$).

\subsection{Non-adiabatic collapse for $\varepsilon^{\rm f}_{\rm dd}>0$}
We begin with a stable condensate with
$\varepsilon^0_{\mathrm{dd}}=0.8$ and suddenly switch to
$\varepsilon^{\rm f}_{\rm dd}>1$.

\subsubsection{Cigar trap}

\begin{figure}[b]
\centering
\includegraphics[width=8.25cm,clip=true]{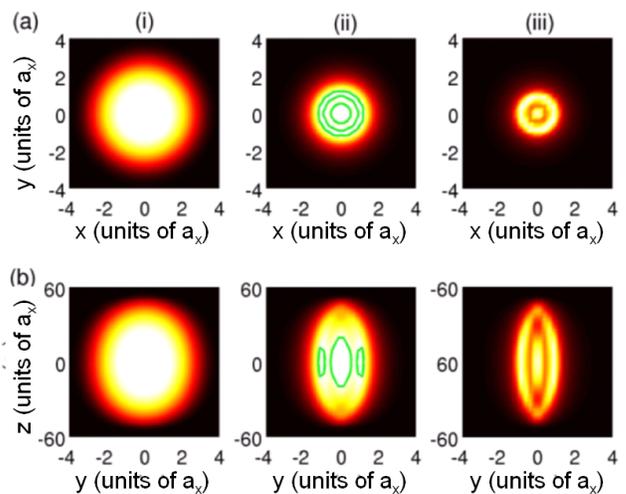}
\caption{(Color online) Snapshots of the collapse of the cigar BEC
with $k_{\rm dd}=2000$ in the (a) {\it x-y} plane and (b) {\it y-z}
plane at (i) $t=0$, (ii) $t=0.47$ and (iii) $t=0.57\omega_x^{-1}$.}
\label{cigar_evo2}
\end{figure}

For a cigar trap $\gamma=0.1$ the initial state lies in the energy
landscape of Fig.~\ref{cigar_evo1}(a). For $t>0$, we switch to
$\varepsilon^{\rm f}_{\rm dd}=1.4$ and an energy landscape of
Fig.~\ref{cigar_evo1}(b).  This system is unstable and the BEC
undergoes a trajectory (green/grey line) towards the collapse
($R_x=0$) region.  According to the TF scaling parameters
$\lambda_x(t)$ and $\lambda_z(t)$  (solid black lines in
Fig.~\ref{cigar_evo1}(c) and (d)), the condensate accelerates to
zero width in the {\it x}-direction after $t\approx 0.75
\omega_x^{-1}$, during which the {\it z}-width reduces by a very
small amount, less than $1\%$. The collapse is therefore highly
anisotropic.  Note that this collapse occurs relatively fast since
the condensate is initially in an elongated state, close to the
collapse threshold.  As we will see the time for collapse is
strongly dependent on the initial shape of the condensate and
therefore on the trap ratio, with more pancake condensates taking
longer to collapse.

In our example the BEC is highly elongated both initially and
throughout its dynamics. Assuming $\kappa(t) \ll 1$ analytic results
for the TF equations of motion can then be obtained. Expanding
$f(\kappa)$ as,
\begin{equation}
\frac{f(\kappa)}{1-\kappa^2}=1+4\kappa^2+3\kappa^2 \log
(\kappa/2)+\mathcal{O}( \kappa^4),
\end{equation}
then Eqs.~(\ref{eq:rxmotion}) and
(\ref{eq:rzmotion}) become, to lowest order,
\begin{eqnarray}
\ddot{\lambda}_{x} \approx -\omega_x^2 \lambda_{x}+ [1- \varepsilon_{\mathrm{dd}}(t)]  \frac{\eta \kappa_{0}g(t)}{\lambda_{x}^{3}\lambda_{z}} \label{eq:lamxeqn} \\
\ddot{\lambda}_{z} \approx -\omega_x^2 \gamma^{2}\lambda_{z}+[1-
\varepsilon_{\mathrm{dd}}(t)] \frac{\eta \kappa_0^3
g(t)}{\lambda_{x}^{2}\lambda_{z}^{2}} \label{eq:lamzeqn}
\end{eqnarray}
Recall that $\eta=15N/4\pi m R_{x0}^5$. Analogous equations have
been derived for an expanding repulsive {\it s}-wave BEC
\cite{castin}. For our time-dependent protocol (\ref{eq:time}) and
to lowest order in $\kappa_0$, Eq.~(\ref{eq:lamzeqn}), has the
solution,
 \begin{eqnarray}
\lambda_{z}(t)& = &\cos (\omega_x \gamma t). \label{eq:lamzsol}
\end{eqnarray}
This corresponds to the limit of the non-interacting gas, and the next correction is of order $\kappa_{0}^{2}$.  To zeroth order in $\gamma$, $\lambda_{z}(t)=1$, and Eq.~(\ref{eq:lamxeqn}) has the solution,
\begin{eqnarray}
\lambda_{x}(t) & = & \frac{1}{\sqrt{2}}\sqrt{(1+\sigma) \cos (2 \omega_x t)+1-\sigma} \label{eq:lamxsol}
\end{eqnarray}
where
\begin{equation}
\sigma=\frac{\eta g_{0}\kappa_0 \edd^0
}{\omega_x^2}(1-\frac{1}{\varepsilon^{\rm f}_{\mathrm{dd}}})
\label{eq:sigmadefn}
\end{equation}
These simplified analytic solutions can give a remarkably good
description of the dynamics. For example, in
Fig.~\ref{cigar_evo1}(c), their prediction of $\lambda_x(t)$ is in
excellent agreement with the full TF equations of motion. For
$\lambda_z(t)$ (Fig.~\ref{cigar_evo1}(d)) deviations are clearly
visible, although the dynamics are very slow in this direction.

We have performed GPE simulations of the collapse for a BEC with
$k_{\rm dd}=80$.  These results (grey dotted lines in
Fig.~\ref{cigar_evo1}(c) and (d)) are not in good agreement with the
TF predictions.  This is because the TF condition for an elongated
dipolar BEC (\ref{eq:cigarTF}) is not remotely satisfied
\cite{Parker08}.  While the BEC approximates a TF profile in the
$z$-direction the transverse profile is more akin to a
non-interacting gaussian ground state.  Under a much larger
interaction strength of $k_{\rm dd}=2000$ (black dotted line), which
does satisfy the TF criterion (\ref{eq:cigarTF}), we find good
agreement with the TF predictions up until $t\approx 0.65
\omega_x^{-1}$.  We have evaluated the energy contributions to the
GPE as outlined in Appendix B and plotted them in
Fig.~\ref{cigar_evo1}(e) for both $k_{\rm dd}=80$ (grey lines) and
$2000$ (black lines).  The validity of the TF approximation for
$k_{\rm dd}=2000$ is confirmed by the smallness of the zero-point
energy in comparison to the other energy contributions.  Indeed, it
remains small up until $t\approx 0.7\omega_x$, thereby validating
the use of the TF approach.  At $t=0$ the dipolar energy $E_{\rm d}$
is negative, indicating the attractive configuration of dipoles in
the initial state. The dipolar energy remains negative and grows in
magnitude as collapse proceeds.  Meanwhile the total kinetic energy
grows and diverges at $t\approx 0.7\omega_x^{-1}$.  The point at
which the energies diverge effectively marks the breakdown of the
validity of the numerical simulations and the TF approach.

In Fig.~\ref{cigar_evo2} we present snapshots from the GPE
simulations with $k_{\rm dd}=2000$. The initial density
[Fig.~\ref{cigar_evo2}(i)] is highly elongated along $z$ and
approximates the TF inverted parabola.  Up to $t \approx 0.4
\omega_x^{-1}$ the BEC collapses anisotropically while maintaining
the inverted parabola shape.  However, from this point in time
[Fig.~\ref{cigar_evo2}(ii)] a local density structure emerges
(highlighted by contours), characterised by modulations in the
density.  This structure evolves to form a striking arrangement of
ellipsoidal shells of high and low density
[Fig.~\ref{cigar_evo2}(iii)], with a high density region in the
centre of the condensate.  The regions of high density grow in
population and peak density, and thereby undergo collapse.  Note
that the development of the density wave structure marks a clear
deviation of the system from the parabolic TF solutions.

According to the Bogoliubov spectrum of a homogeneous system (to be
discussed further in Section \ref{sec:length_time}) the dipolar BEC
can be dynamically unstable to phonon modes \cite{goral,eberlein05}.
This instability has a strong dependence on angle relative to the
polarization direction, and physically this represents the tendency
for the system to preferentially align the dipoles in an end-to-end
configuration in the polarization direction.  This effect gives rise
to the highly elongated shell structures observed here.

\subsubsection{Pancake trap}
\begin{figure}[t]
\centering
\includegraphics[width=8.25cm,clip=true]{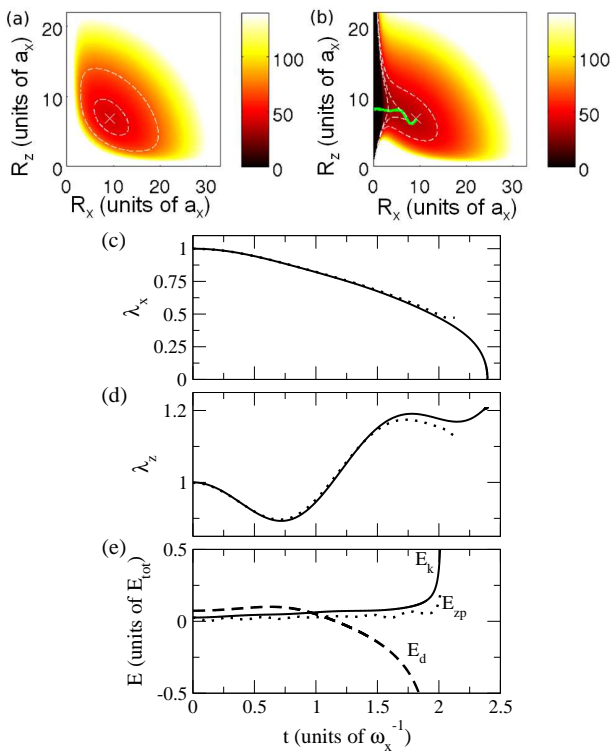}
\includegraphics[width=5.5cm,clip=true]{Fig4b.eps}
\caption{(Color online) Collapse dynamics in a pancake trap
$\gamma=2$ for $\varepsilon^0_{\rm dd}=0.8$ and $\varepsilon^{\rm
f}_{\rm dd}=1.4$. (a) Energy landscape of Eq.~(\ref{eq:E}) at $t=0$.
(b) Energy landscape for $t>0$, with the TF trajectory indicated
(green/grey line).  Energy is scaled in units of $\hbar\omega_x$.
(c) $\lambda_x(t)$ from the TF equations of motion (solid line) and
the GPE with $k_{\rm dd}=80$ (dotted line). (d) Same for
$\lambda_z(t)$.  (e) Kinetic $E_{\rm k}$, zero-point $E_{\rm zp}$
and dipolar $E_{\rm d}$ energies from the GPE, renormalised by the
total energy $E_{\rm tot}$.} \label{pancake_evo1}
\end{figure}

\begin{figure}[t]
\centering
\includegraphics[width=8cm,clip=true]{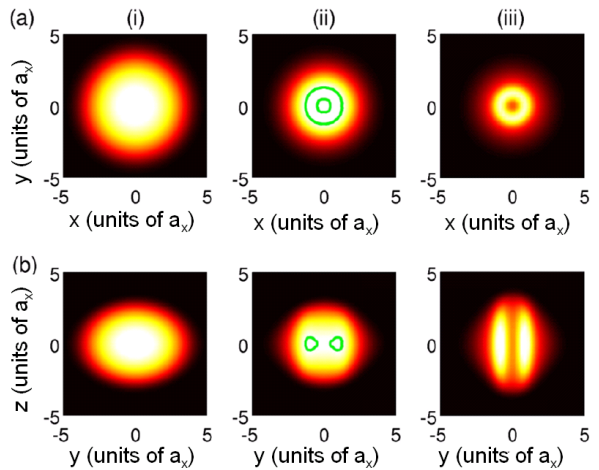}
\caption{(Color online) Snapshots of the BEC density during the
collapse of Fig.~\ref{pancake_evo1} in (a) {\it x-y} plane and (b)
{\it y-z} plane at (i) $t=0$, (ii) $t=1\omega_x^{-1}$ and (iii)
$t=1.5 \omega_x^{-1}$.  Light/dark regions correspond to high/low
density, while contours in (ii) help to visualise the density
structures.  } \label{pancake_frames1}
\end{figure}
We now consider a BEC within a pancake trap $\gamma=2$.  The initial
state with $\varepsilon^0_{\rm dd}=0.8$ resides in the energy
landscape of Fig.~\ref{pancake_evo1}(a). Following the sudden switch
to $\varepsilon^{\rm f}_{\rm dd}=1.4$ and an energy landscape of
Fig.~\ref{pancake_evo1}(b), the BEC follows a trajectory (green/grey
line) towards the collapse $R_x=0$ region.  According to the TF
equations of motion (solid lines in Fig.~\ref{pancake_evo1}(c) and
(d)), the condensate accelerates to zero width in the {\it
x}-direction after $t\approx 2 \omega_x^{-1}$, while the {\it
z}-width oscillates by approximately $10\%$. Again, the collapse is
highly anisotropic.  It is considerably slower than in the cigar
trap because the initial condensate is in a more stable state,
dominated by repulsive interactions.  Note that if the condensate
begins in a highly flattened state $\kappa\gg 1$, it elongates over
time and its aspect ratio will be reversed.  As such no expansion of
$\kappa$, analogous to Eqs.~(\ref{eq:lamxeqn}) and
(\ref{eq:lamzeqn}), is appropriate.

For $k_{\rm dd}=80$ the TF pancake criterion (\ref{eq:pancakeTF}) is
satisfied. The corresponding GPE predictions (dotted lines in
Fig.~\ref{pancake_evo1}(c) and (d))
agree well with the TF predictions up to $t \approx 2\omega_x^{-1}$,
with the zero-point energy remaining small up until this point.
Initially, the dipolar energy is positive due to the dominance of
repulsive dipoles in the initial flattened BEC but as the BEC
collapses, it becomes negative and diverges. Again, collapse causes
a divergence in the kinetic energy.

\begin{figure}[t]
\centering
\includegraphics[width=6.cm,clip=true]{Fig6a.eps}
\includegraphics[width=8cm,clip=true]{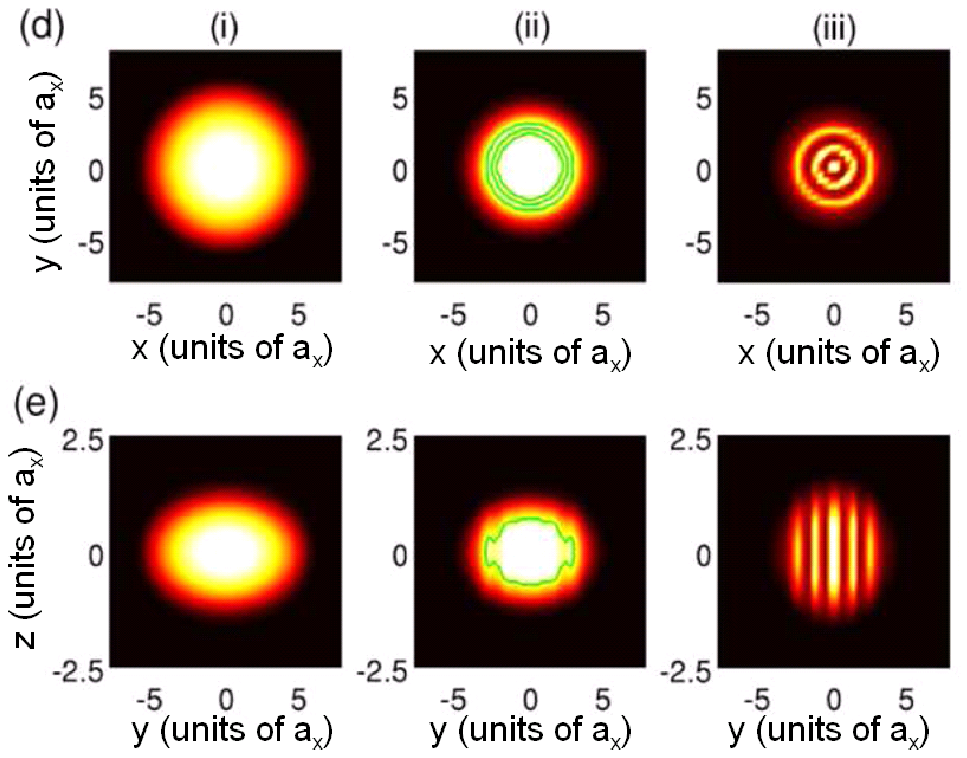}
\caption{(Color online) Dynamics in a pancake trap $\gamma=5$ under
a sudden change from $\varepsilon_{\rm dd}=0.8$ to $4$. (a)
$\lambda_x(t)$ and (b) $\lambda_z(t)$ from the TF equations (solid
line) and the GPE with $k_{\rm dd}=80$ (dotted line). (c) Kinetic
$E_{\rm k}$, zero-point $E_{\rm zp}$ and dipolar $E_{\rm d}$
energies from the GPE simulations. (d)-(e) Density snapshots during
the GPE simulations in the (d) {\it x-y} plane and (e) {\it y-z}
plane at (i) $t=0$, (ii) $1$ and (iii) $1.5\omega_x^{-1}$.}
\label{pancake_frames2}
\end{figure}

During the collapse we again see the formation of a local density
structure, as shown in Fig.~\ref{pancake_frames1}(a,b). These
structures are not closed ellipsoidal shells but now rectilinear
shells aligned in the $z$-direction.  In this case it is likely that
the large trap frequency in $z$, which leads to a large excitation
energy for axial excitations, suppresses the curvature of the
shells. Here the structure features a node of low density for $r=0$
and a single high density shell at finite radius.  This structure
evolves rapidly with atoms moving away from the centre of the
condensate to populate the outer shell. This causes the observed
flattening in $\lambda_x$ at $t \approx 2\omega_x^{-1}$. By later
times [Fig.~\ref{pancake_frames1}(iii)] the shell has elongated
axially and shrunk radially, and the system continues to undergo
local collapse.

To probe the effect of the trap geometry we also present the
collapse dynamics for $\gamma=5$.  For $\varepsilon^{\rm f}_{\rm
dd}=1.4$ no collapse occurs and so we employ a more extreme value of
$\varepsilon^{\rm f}_{\rm dd}=4$.  The TF equations of motion (solid
lines in Fig.~\ref{pancake_frames2}(a) and (b)) predict that the
condensate undergoes shape oscillations rather than collapse.  The
GPE results (dotted lines) agree up to $t=1.5\omega_x^{-1}$.  During
this time the condensate density
(Fig.~\ref{pancake_frames2}(d-e)(i)) remains approximately an
inverted parabola, although weak local density perturbations begin
to emerge (Fig.~\ref{pancake_frames2}(d-e)(ii)).  However, for
$t>1.5\omega_x^{-1}$, the kinetic energy $E_{\rm k}$ diverges to
$+\infty$ and $E_{\rm d}$ diverges to $-\infty$.   We again observe
the development (and collapse) of local density structure
(Fig.~\ref{pancake_frames2}(d-e)(iii)). This structure features
considerably more shells than our previous example, primarily due to
the larger radial extent of the condensate.  It also features a
central density anti-node, rather than a node.

\subsection{Non-adiabatic collapse for $\varepsilon^{\rm f}_{\rm dd}<0$}
Here we start with a stable condensate with $\varepsilon^0_{\rm
dd}=-0.2$ and suddenly switch to $\varepsilon^{\rm f}_{\rm
dd}<-0.5$.

\subsubsection{Cigar trap}
\begin{figure}[b]
\centering
\includegraphics[width=6cm,clip=true]{Fig7a.eps}\\
\includegraphics[width=7.75cm,clip=true]{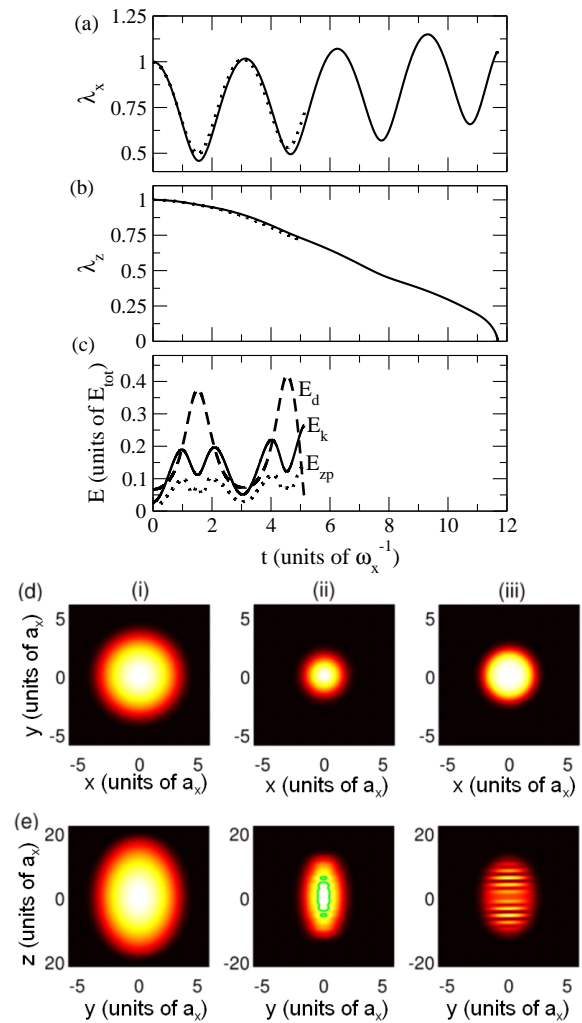}
\caption{(Color online) Collapse dynamics in a cigar trap
$\gamma=0.2$ for $\varepsilon^0_{\rm dd}=-0.2$ and $\varepsilon^{\rm
f}_{\rm dd}=-4$. (a) $\lambda_x(t)$ and (b) $\lambda_z(t)$ from the
TF equations of motion (solid line) and the GPE with $k_{\rm dd}=80$
(dotted line).(c) Kinetic $E_{\rm k}$, zero-point $E_{\rm zp}$ and
dipolar $E_{\rm d}$ energies.  (d)-(e) GPE density in the (d) {\it
x-y} plane and (e) {\it y-z} plane at (i) $t=0$, (ii) $4.4$ and
(iii) $5\omega_x^{-1}$.} \label{cigar_neg}
\end{figure}

In an elongated trap $\gamma=0.2$ we have observed collapse for
$\varepsilon^f_{\rm dd}=-4$ with the dynamics presented in
Fig.~\ref{cigar_neg}.  Under the TF equations of motion (solid
lines), the condensate undergoes large oscillations in the {\it
x}-direction and collapses slowly in {\it z}.  The collapse occurs
predominantly in the $z$ direction towards an infinitely thin plane
of side-by-side dipoles.  This is the opposite to the regime of
$\edd>0$.  Note that the collapse is slow because the condensate is
initially in an elongated state where the attractive interactions
that induce collapse are weak.
With $k_{\rm dd}=80$ the GPE initial state is in the TF pancake
regime (\ref{eq:cigarTF}). The GPE results agree well up to their
point of validity at $t\approx 5\omega_x^{-1}$.  During this time,
the zero-point kinetic energy remains relatively small, and the
kinetic and dipolar energies undergo large oscillations due to the
radial shape oscillations.  At $t\approx5\omega_x^{-1}$, $E_{\rm k}$
and $E_{\rm d}$ diverges signifying the limit of validity of the
simulations.

Consideration of the condensate density profile during collapse reveals that the condensate develops weak planar density corrugations by $t\approx4\omega_x^{-1}$ [Fig.~\ref{cigar_neg}(d,e)(ii)].  These become amplified into a striking planar density perturbation of approximately $10$ localised planes of dipoles, aligned in the $x-y$ plane [Fig.~\ref{cigar_neg}(d,e)(iii)].  This is the same phenomenon as observed earlier but with perpendicular orientation due to the fact that the $\edd<0$ dipoles are now attractive when side-by-side.

\subsubsection{Pancake trap}
\begin{figure}[t]
\centering
\includegraphics[width=6cm,clip=true]{Fig8a.eps}\\
\includegraphics[width=7.75cm,clip=true]{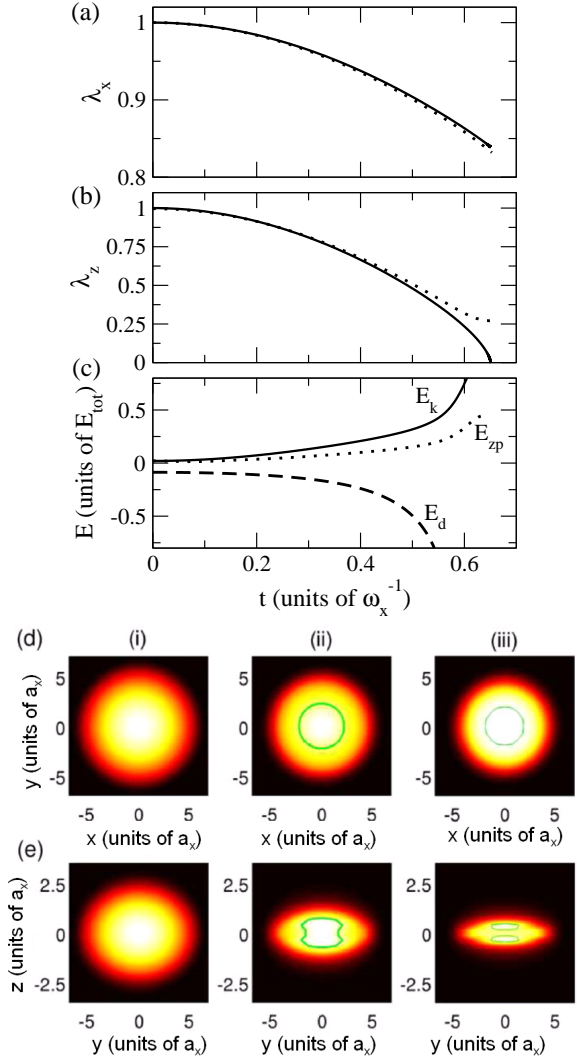}
\caption{(Color online) Collapse dynamics in a pancake trap
$\gamma=2$ under a sudden change from $\varepsilon_{\rm dd}=-0.2$ to
$-0.8$. (a) $\lambda_x(t)$ and (b) $\lambda_z(t)$ according to the
full TF equations of motion (solid line) and GPE simulations with
$k_{\rm dd}=80$ (dotted line).(c) Kinetic $E_{\rm k}$, zero-point
$E_{\rm zp}$ and dipolar $E_{\rm d}$ energies from the GPE,
renormalised by the total energy $E_{\rm tot}$. (d)-(e) Density
snapshots in the (d) {\it x-y} and (e) {\it y-z} planes at (i)
$t=0$, (ii) $0.42$ and (iii) $0.53\omega_x^{-1}$. }
\label{pancake_neg}
\end{figure}

For $\varepsilon^{\rm f}_{\rm dd}=-0.8$, the condensate dynamics within a pancake trap $\gamma=2$ are presented in Fig.~\ref{pancake_neg}.  According to the TF equations of motion, $\lambda_z(t)$ accelerates to zero within $t\approx0.6\omega_x^{-1}$ while $\lambda_x(t)$ decreases more slowly.
Taking $k_{\rm dd}=80$ the initial ground state satisfies the pancake TF criteria (\ref{eq:pancakeTF}) and we see excellent agreement with the TF predictions up to $t\approx0.6\omega_x^{-1}$, during which $E_{\rm zp}$ remains relatively small.  However, beyond this point $E_{\rm zp}$ diverges, as does $E_{\rm k}$ and $E_{\rm d}$.

Like the cigar case, the condensate density develop planar density perturbations, as shown in Fig.~\ref{pancake_neg}(d,e)(ii) and (iii).  However, given that the pancake system is narrower in $z$, only two planes of high density emerge.

\section{General properties of collapse \label{sec:length_time}}
We now map out the general behaviour of the important length scales and time scales of the collapse, and the critical trap ratio that can stabilise against collapse.

\subsection{Length scales for collapse {\label{sec:length}}}
\begin{figure}[t]
\centering
\includegraphics[width=8.4cm,clip=true]{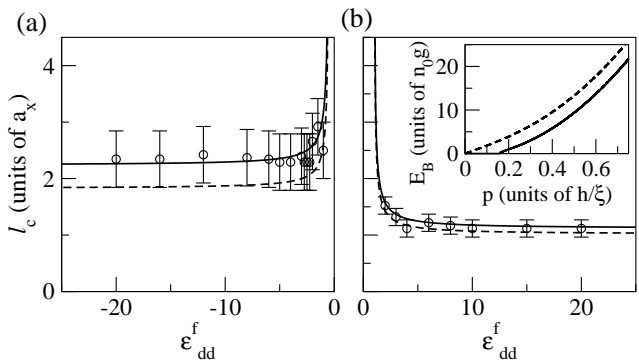}
\caption{Length scale of collapse $l_{\rm c}$ as a function of
$\varepsilon^{\rm f}_{\rm dd}$ according to GPE simulations
(circles) and the predictions of Eq.~(\ref{eqn:roton1}) and
(\ref{eqn:roton2}) for a homogeneous system $A=B=1$ (dashed lines)
and for a trapped system $A \neq B \neq 1$ (solid line).  (a) The
regime $\edd<-0.5$, assuming a cigar-shaped BEC with $\gamma=0.2$,
$\varepsilon^0_{\rm dd}=-0.2$ and $k_{\rm dd}=-80$.  For the cigar
prediction (solid line) we assume a TF transverse profile for which
$B=\sqrt{3/2}$.  (b) The regime $\edd>1$, assuming a pancake
geometry $\gamma=5$ with $\varepsilon^0_{\rm dd}=0.8$ and $k_{\rm
dd}=80$.  For the pancake prediction (solid line) we assume a TF
axial profile for which $A=\sqrt{5/4}$.  Inset: Homogeneous
Bogoluibov spectrum of Eq.~(\ref{eq:bogdispersion}) for $\edd=1.4$
and for $\theta=0$ (dotted line) and $\theta=\pi/2$ (dot-dashed
line).  Note that the error bars in the GPE results arise from the
grid discretization.} \label{fig:length}
\end{figure}

A homogenous dipolar BEC is unstable to periodic density
perturbations (phonons) when $\varepsilon_{\mathrm{dd}}>1$ or
$\edd<-0.5$ \cite{goral}. This can be seen immediately from the
Bogoliubov dispersion relation between the energy $E_{\mathrm{B}}$
and momentum $p$ for phonons in the gas, given by,
\begin{equation}
E_{\mathrm{B}}=\sqrt{\left(\frac{p^{2}}{2m}\right)^{2}
+2gn\left\{1+\varepsilon_{\mathrm{dd}}
\left(3 \cos^{2} \theta -1 \right) \right\}\frac{p^{2}}{2m}}.
\label{eq:bogdispersion}
\end{equation}
A mode with energy $E_{\rm B}$ evolves as $\exp(iE_{\rm B}t/\hbar)$
and so when $E_{\rm B}$ becomes imaginary, the mode grows
exponentially, i.e., a dynamical instability. This dispersion
relation depends on $\theta$, the angle between the momentum of the
phonon and the external polarizing field.  For $\edd>1$, phonons
propagating perpendicularly to the polarization direction
($\theta=\pi/2$) undergo a dynamical instability while no
instability occurs along the polarization direction ($\theta=0$).
This is illustrated in Fig.~\ref{fig:length}(b) (inset) which plots
the spectrum (\ref{eq:bogdispersion}) for $\edd=1.4$ and for the
polarizations $\theta=0$ (dashed line) and $\theta=\pi/2$ (solid
line); the point where the solid line touches zero marks the
transition to instability.  In an infinite and initially homogeneous
system we expect this instability to break the condensate up into a
lattice of filaments, i.e.\ cylindrical structures aligned along the
polarization direction.   In trapped condensates, shells or planes
(aligned along $z$) may be favored instead.  Meanwhile, when
$\edd<-0.5$ (we remind the reader that we mean in this case that
$C_{\rm dd}<0$),  it is phonons propagating along the polarization
direction ($\theta=0$) that undergo a dynamical instability while no
instability occurs in the perpendicular direction ($\theta=\pi/2$).
This suggests an instability towards stratification into planes
lying perpendicular to the polarization axis.  These predictions are
consistent with our observations in the previous section.

We can estimate the characteristic length scale of the collapse
structures from the homogeneous Bogoliubov spectrum
(\ref{eq:bogdispersion}).  If $p_{c}$ is the critical momentum for
which the dispersion relation passes through zero energy (which
signifies the onset of dynamical instability), then we can expect
the characteristic length scale to be $l_{c}=2 \pi \hbar/ p_{c} $.
Although we are concerned with trapped, inhomogeneous condensates,
the same length scale $l_{c}$ will apply providing the system is of
size $R \gg l_{c}$. From Eq.~(\ref{eq:bogdispersion}) it is straight
forward to show that this length scale is given by,
\begin{eqnarray}
l_c&=& A \frac{\sqrt{2}\pi\xi}{\sqrt{\varepsilon_{\rm
dd}-1}}~~~~~~~{\rm for}~~~\edd>1
\label{eq:lengthscales1a} \\
l_c&=&B \frac{\sqrt{2}\pi\xi}{\sqrt{|1+2\varepsilon_{\rm dd}|}}~~~{\rm for}~~~\edd<-0.5.
\label{eq:lengthscales1}
\end{eqnarray}
Here $\xi=1/\sqrt{8\pi n_0 |a_{\rm s}|}$ is the {\it s}-wave healing
length at the condensate centre. $A$ and $B$ are factors that take
account of the trapping, and are unity in a homogenous system.

We can improve the applicability of our length scale predictions to
inhomogeneous systems by taking into account the trapping in one or
two directions. In an infinite cylindrical BEC with tight radial
trapping, such that the radial density profile is a gaussian, it is
known that the axial speed of sound is reduced by a factor of
$\sqrt{2}$ in comparison to a uniform system with the same peak
density $n_0$ \cite{zaremba98}. This rescaling of the dispersion
relation arises because the average density is in fact given by
$n_{0}/2$. In general, we can obtain the average density for both
cigar and pancake condensates by integrating out the tightly
confined direction.  We will perform this for the relevant cases of
gaussian and parabolic (TF) density profiles. The resulting
modification to the dispersion relations along the weakly confined
direction(s) gives the trapping parameters $A$ and $B$
which occur in Eq.~(\ref{eq:lengthscales1a}) and
(\ref{eq:lengthscales1}). For $\varepsilon_{\rm dd}>1$ the
interesting case is that of a pancake, for which $A \rightarrow
A_{\mathrm{Gauss}}=2^{1/4}\approx 1.2$ in the gaussian case and
$A \rightarrow A_{\mathrm{TF}}=\sqrt{5/4}\approx 1.1$ in
the TF case.  For $\varepsilon_{\rm dd}<-0.5$ the interesting case
is that of a cigar, for which $B \rightarrow
B_{\mathrm{Gauss}}=\sqrt{2}\approx 1.4$ in the gaussian case and
$B \rightarrow B_{\mathrm{TF}}=\sqrt{3/2}\approx 1.2$ in the
TF case.

In order to relate our numerical results on adiabatic and
non-adiabatic collapse to the unstable modes of
Eq.~(\ref{eq:bogdispersion}), let us consider the case of positive
$\edd$.  For $\edd=1$ the collapse length scale is infinite. If
$\edd$ is increased adiabatically then $l_c$ decreases. At some
critical $\edd$, $l_c$ becomes equal to the condensate size $R$ and
a dynamical instability occurs on the length scale of $R$. Thus,
according to Eq.\ (\ref{eq:bogdispersion}), if the point of collapse
is reached adiabatically, global collapse will occur.  This picture
is qualitatively consistent with our observations in Section
\ref{sec:solutions} in the range $-1\ltsimeq \edd\ltsimeq2$, where
we observed that adiabatic collapse proceeds in a global manner.
However, outside of this regime we saw evidence for adiabatic local
collapse, which we attributed to the presence of a roton minimum in
the excitation spectrum. The homogeneous dispersion relation
(\ref{eq:bogdispersion}) does not have a roton minimum and in order
to introduce one it is necessary to explicitly include a trap in at
least one dimension. In highly cigar-shaped or pancake-shaped
systems the roton minimum occurs in the dispersion relation for
low-energy excitations along the weakly confined direction. The
momentum at which the roton minimum occurs, $p_{r}$, defines a
length scale $\l_{r}=2 \pi \hbar/p_{r}$ which is usually closely
related the system size in the tightly trapped direction
\cite{fischer06,ronen07}. Adiabatic local collapse is therefore
expected to take place on length scales $l_{r}$ where the roton
minimum touches the zero energy axis. We have compared simple
analytic predictions for the dispersion relations of an infinite
cigar \cite{giovanazzi04a} and an infinite pancake \cite{fischer06},
which do include a roton minimum, with the homogeneous result
(\ref{eq:bogdispersion}).  We find that the value of $l_c$ can
significantly differ from the predictions of
(\ref{eq:lengthscales1a}) and (\ref{eq:lengthscales1}) when the
system is close to the collapse threshold, but quickly become very
similar as we move deeper into the collapse regime. Thus, we expect
that an adiabatic collapse experiment could provide clear evidence
for the presence of a roton minimum in the excitation spectrum, but
a non-adiabatic collapse experiment would be less conclusive. To
tackle this problem properly one should numerically obtain the full
Bogoliubov spectrum of a trapped system \cite{ronen}.

We now consider the situation where we suddenly switch from
$\edd^0<1$ to $\edd^{\rm f}\gg1$. Here we go from the regime where
$l_c \gg R$ to $l_c<R$, and a dynamical instability occurs on a
local scale.  Although the dynamical instability evolves for all
length scales in the range $l>l_c$, the imaginary energy eigenvalue
is largest for $l=l_c$ and this unstable mode dominates the system.
This is consistent with our observations in Section
\ref{sec:collapse} where, following a sudden change in $\edd$,
collapse evolves mainly through local density structures.

We now specifically consider the case where collapse is induced
suddenly by modifying the {\it s}-wave interactions, while the
dipolar interactions remains constant.  Then we can rewrite
Eqs.~(\ref{eq:lengthscales1a}) and (\ref{eq:lengthscales1}) as,
\begin{eqnarray}
l_c&=&A \sqrt{\frac{\pi \varepsilon^{\rm f}_{\rm dd}}{4 n_0 a_{\rm dd}(\varepsilon^{\rm f}_{\rm dd}-1)}}~~~~~~~{\rm for}~~~\edd>1 \label{eqn:roton1} \\
l_c&=&B \sqrt{\frac{\pi \varepsilon^{\rm f}_{\rm dd}}{4 n_0 |a_{\rm dd}(1+2\varepsilon^{\rm f}_{\rm dd})|}}~~~~{\rm for}~~~\edd<-0.5
\label{eqn:roton2}
\end{eqnarray}

These predictions are plotted in Fig.~\ref{fig:length}(a) and (b),
for the regimes of $\edd<-0.5$ and $\edd>1$, respectively.  We have
conducted a series of GPE simulations for trapped dipolar BECs to
determine the true collapse length scale (defined as the distance
between peaks in adjacent shells), with these predictions being
shown by the circles.  Note that in Fig.~\ref{fig:length}(a) we
employ a cigar-shaped BEC with $\gamma=0.2$ and $\varepsilon^0_{\rm
dd}=-0.2$, and in Fig.~\ref{fig:length}(b) we employ a
pancake-shaped BEC with $\gamma=5$ and $\varepsilon^0_{\rm dd}=0.8$.
Even in the homogeneous limit $A=B=1$ (dashed lines) the analytic
predictions for $l_c$ are in very good agreement with the
simulations. Examination of the GPE solutions reveals that the
density profile in the tightly confined direction is closely
approximated by the TF profile.  The analytic results using the
appropriate trapping parameters ($A=\sqrt{5/4}$ and $B=\sqrt{3/2}$)
are shown by solid lines in Fig.~\ref{fig:length}.
 With these trapping parameters included the agreement becomes excellent,
and clearly demonstrates the importance of taking the trapping into account.

\subsection{Time for global collapse}
We now make some simple predictions for the characteristic collapse
time $\tau_c$.  Having demonstrated that the TF model gives a good
description of collapse we will employ it exclusively here. Since
the TF equations of motion cannot describe local collapse, our
analysis is limited to global collapse. Furthermore we will only
consider non-adiabatic collapse since the time scale for adiabatic
collapse is technically infinite.

Experimentally, the collapse time is the time at which a `sudden'
depletion of the condensate occurs due to three-body loss
\cite{lahaye}.  The suddenness arises because of the dramatic
scaling of losses with time: the rate of three-body loss scales as
$n^3$, and the density $n$ itself accelerates in amplitude during
collapse. Consequently, a good estimate for $\tau_c$ is the time
over which the peak density diverges or, equivalently, the time over
which one or more radii tend to zero \cite{ticknor}.

\begin{figure}[b]
\centering
\includegraphics[width=8.5cm,clip=true]{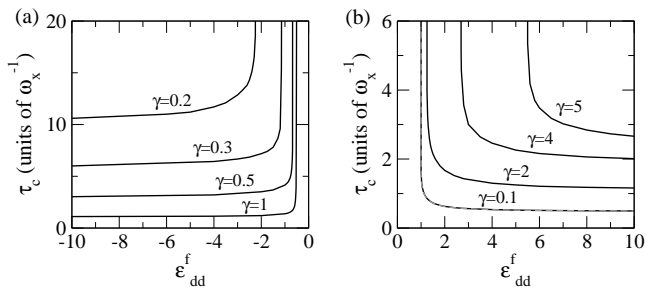}
\caption{Collapse time $\tau_{\rm c}$ following a sudden change in
$\edd$ according to the TF equations of motion.  Various trap ratios
are presented. (a) $\edd^0=-0.2$ and $\edd^{\rm f}<0$. (b)
$\edd^0=0.8$ and $\edd^{\rm f}>0$. For the case of $\gamma=0.1$ we
also plot the analytic expression of Eq.~(\ref{eq:tauc}) (grey
dashed line).} \label{collapse_times}
\end{figure}

The TF collapse time can be obtained, in general, by numerical
solution of the TF equations of motion.  However, for highly
elongated BECs, Eqs.~(\ref{eq:lamzsol}) leads to an analytic form
for $\tau_c$ given by,
\begin{equation}
\tau_{c}=\frac{1}{2} \arccos \left[
\frac{\sigma-1}{\sigma+1}\right], \label{eq:tauc}
\end{equation}
where $\sigma$ is given by Eq.~(\ref{eq:sigmadefn}).  It should be
noted that in the limit $\varepsilon^{\rm f}_{\mathrm{dd}}
\rightarrow \infty$, Eq.~(\ref{eq:sigmadefn}) becomes,
\begin{equation}
\lim_{\varepsilon^{\rm f}_{\mathrm{dd}}\to\infty} \sigma = \edd^0
\frac{\eta g_{0} \kappa_{0}}{\omega_{x}^{2}},
\end{equation}
where we recall that $\eta=15N/4\pi m R_{x0}^5$. Thus the limiting
value of $\tau_c$ is determined only by the initial parameters
$\gamma$ and $\varepsilon^0_{\mathrm{dd}}$.

In Fig.~\ref{collapse_times}(a) and (b) we show how $\tau_c$ depends
on the final interaction parameter $\varepsilon^{\rm
f}_{\mathrm{dd}}$ for initial values of $\varepsilon^0_{\rm
dd}=-0.2$ and $0.8$, respectively.  For weak interactions $\tau_c$
diverges as the interactions become too weak to induce collapse,
while in the limit of large interactions (positive or negative
$\edd$), $\tau_c$ tends towards a finite value, as expected.    For
$\edd<0$, elongated systems become more stable and are the slowest
to collapse.  In contrast, for $\edd>0$, elongated systems are least
stable and therefore the fastest to collapse.
\begin{figure}[t]
\centering
\includegraphics[width=8.5cm,clip=true]{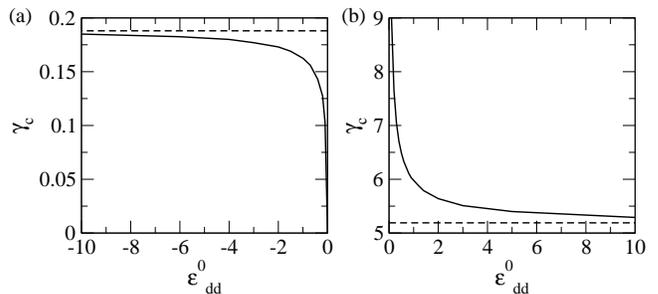}
\caption{ Critical trap ratio $\gamma_{\rm c}$ that stabilises the
BEC against collapse as a function of initial interaction strength
$\varepsilon^0_{\rm dd}$, according to the full TF equations of
motion. The final interaction strength $\varepsilon^{\rm f}_{\rm dd}=10^4$
is so large that it is effectively the infinite limit.}
\label{crit_ratio}
\end{figure}
Note that for the highly elongated case of $\gamma=0.1$, the
analytic collapse time of Eq.~(\ref{eq:tauc}) is in excellent
agreement with the full TF equations of motion.

\subsection{Critical trap ratio for global collapse}
In the range of $\edd$ considered in this work, there exists a
critical trap ratio $\gamma_c$ which can stabilise the parabolic TF
solutions against {\em global} collapse.  As we showed in Section
III, if collapse is approached {\em adiabatically} this critical
trap ratio is fixed, being $\gamma_c=5.2$ for positive $\edd$ and
$\gamma_c=0.19$ for negative $\edd$.  However, if collapse is
induced non-adiabatically, the critical trap ratio becomes a
function of $\varepsilon^0_{\rm dd}$.  To examine this threshold for
global collapse in more detail, Fig.~\ref{crit_ratio} plots the
critical trap ratio $\gamma_c$ as a function of the initial
interaction parameter $\varepsilon^0_{\rm dd}$ for both negative and
positive $\edd$. Note that we have considered $|\varepsilon^{\rm
f}_{\rm dd}|=10^{4}$, which is so large that it effectively behaves
like the infinite limit.  For $\varepsilon^0_{\rm dd}<0$, $\gamma_c$
tends towards zero for $\varepsilon^0_{\rm dd}\rightarrow 0$, and in
the opposing limit of $\varepsilon^0_{\rm dd}\rightarrow -\infty$ it
increases asymptotically towards the adiabatic value
$\gamma_c=0.19$. Conversely, for $\varepsilon^0_{\rm dd}>0$,
$\gamma_c$ diverges as $\varepsilon^0_{\rm dd}\rightarrow 0$, and in
the opposing limit of $\varepsilon^0_{\rm dd}\rightarrow \infty$ it
decreases towards the static value $\gamma_c=5.17$.  Note that,
providing $|\varepsilon^0_{\rm dd}|\gg0$, $\gamma_c$ varies only
weakly with $\varepsilon^0_{\rm dd}$ and becomes very close to the
static value of $\gamma_c$.

\section{Discussion \label{sec:disc}}

In a recent series of experiments the Stuttgart group demonstrated
the collapse of a dipolar condensate \cite{koch,lahaye}. A Feshbach
resonance was employed to give time-dependent control over $a_{\rm
s}$ and therefore $\varepsilon_{\rm dd}$. In Ref. \cite{koch},
$a_{\rm s}$ was reduced slowly (over several trap periods) to probe
the stability of the ground state to collapse.  For $\gamma=1$,
collapse was observed for $\varepsilon_{\rm dd}\approx 1$, which is
in good agreement with the static solutions in
Fig.~\ref{fig:kappa}(a).  For $\gamma \approx 10$, the BEC was
stabilised under purely dipolar interactions, i.e. the limit
$\varepsilon_{\rm dd}\rightarrow \infty$, in good qualitative
agreement with our understanding of the role of trap geometry.
\begin{figure}[t]
\centering
\includegraphics[width=7.75cm,clip=true]{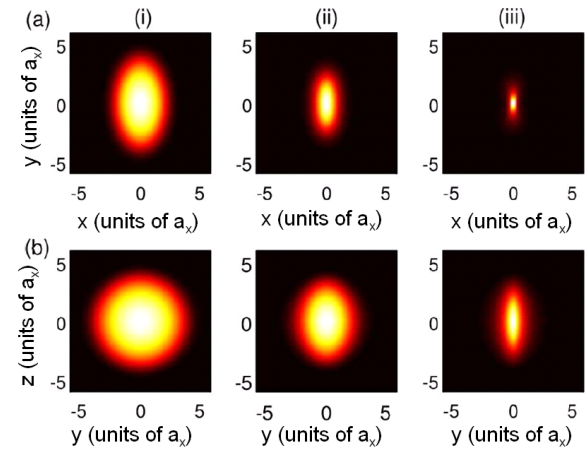}
\caption{(Color online) Collapse dynamics in Stuttgart system under
the linear decrease of $a_{\rm s}$ over time $\tau_r=1$ms. Density
in the (a) {\it x-y} plane and (b) {\it y-z} plane at (i) $t=0$,
(ii) $t=3.1$ and (iii) $t=3.7\omega_x^{-1}$.} \label{pfau_evo1}
\end{figure}

\begin{figure}[b]
\centering
\includegraphics[width=7.75cm,clip=true]{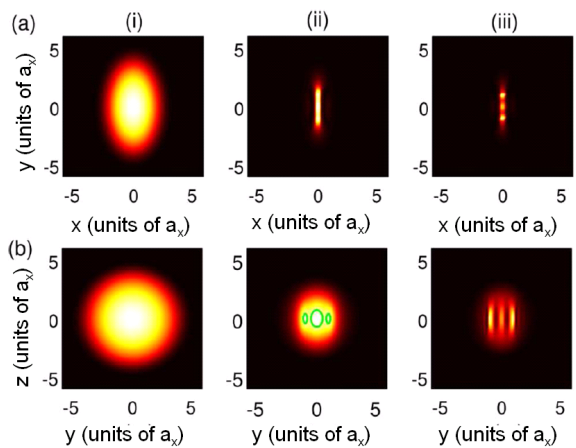}
\caption{(Color online) Collapse dynamics in Stuttgart system under
the sudden decrease in $a_{\rm s}$. Density in the (a) {\it x-y}
plane and (b) {\it y-z} plane at (i) $t=0$, (ii) $t=1.08$ and (iii)
$t=1.16\omega_x^{-1}$.} \label{pfau_evo2}
\end{figure}
In Ref.\ \cite{lahaye}, collapse was observed in an almost spherical
trap.  From an initial BEC at $t=0$ with $a^{\rm 0}_{\rm s}=1.59$nm,
the scattering length was reduced linearly to a final value $a^{\rm
f}_{\rm s}=0.27$nm over a time scale $\tau \sim 1$ms.  The BEC was
observed to collapse anisotropically towards a narrow cigar shape,
aligned in $z$, over a time scale of $\tau^{\rm exp}_{\rm c}\approx
1.5$ms.  We will see that this time scale is sufficiently long that
local collapse is not induced. Following this initial collapse an
explosion occurred, resulting in a spectacular state with a shape
resembling that of a $d$-wave orbital. Numerical simulations of the
GPE with three-body loss were in excellent agreement with the
observed dynamics \cite{lahaye}.

The experimental collapse appears to occur globally with no signs of
local collapse.  To resolve the issue of the apparent absence of local collapse, we have
simulated the experimental dynamics based on $N=20,000$ atoms and a
fully anisotropic trap $(\omega_x, \omega_y,
\omega_z)=2\pi\times(660, 400, 530)$Hz.  Under linear ramping of
$a_{\rm s}$ over time $\tau=1$ms, our results in
Fig.~\ref{pfau_evo1} confirm that the condensate collapses globally
rather than locally.  The condensate collapses, mainly in the $x-y$
plane, and forms a very narrow cigar-shaped BEC.   Because
$\omega_x>\omega_y$, the transverse collapse is quickest in the
$x$-direction.  The elongated collapsed state
(Fig.~\ref{pfau_evo1}(iii)) forms after a collapse time $\tau_{\rm
c}\approx 1$ms.  This is quicker than the time scales reported in
\cite{lahaye}, where it is known that eddy currents suppress the
applied change in $a_{\rm s}$ and effectively extend the ramping
time by a factor of two or three.  This does not affect our
qualitative results but merely slows down the collapse.  Indeed, if
we employ a ramping time of $\tau=2$ms, we find $\tau_{\rm c}\approx
1.6$ms, in agreement with \cite{lahaye}.

The experimental ramping time is far from being sudden since it is
of the order of the trap period that characterises the internal
dynamics of the BEC.  We can therefore expect that global collapse
will be initiated well before any local instabilities. Indeed, if we
make a more sudden change of $a_{\rm s}$, e.g., $\tau=0$, then we
see in Fig.~\ref{pfau_evo2} that local collapse now occurs. However,
within this noncylindrically symmetric geometry we observe the
formation of parallel density stripes, rather than shells. Again,
the $x$-direction collapses towards zero width. However, the
$y$-direction does not shrink globally but develops a corrugated
structure that enables the dipoles to predominantly align along $z$.
These stripes become amplified and collapse themselves.

\section{Conclusions \label{sec:concs}}

We have studied the collapse dynamics of a dipolar Bose-Einstein
condensate, triggered either by an adiabatic or non-adiabatic change
in the dipolar-to-{\it s}-wave ratio $\edd=C_{\rm dd}/3g$. In
general, the collapse occurs anisotropically
 and is driven by the dipoles seeking to line up end-to-end for $\edd>0$ and
  side-by-side for $\edd<0$. In the case of adiabatic collapse, where the
  ground state solutions are followed up until the instant of collapse, we
   observe both global and local collapse. In the range $-1\ltsimeq\edd \ltsimeq 2$
   we find global collapse towards a single line or plane of dipoles. Outside of this
   range we have seen adiabatic local collapse which we suggest is a signature of a
   roton minimum in the excitation spectrum.  Similar theoretical predictions have also recently been reported by Bohn {\it et al.} \cite{bohn}.
   Note that care must be taken to distinguish
   such local collapse from the results of non-adiabatic collapse.

If collapse is triggered non-adiabatically via a sudden change in $\edd$, the instability can jump to length scales much less than the condensate size, resulting in local collapse.  We have analysed this instability over the considerable range $-10\ltsimeq\edd \ltsimeq 20$.  This instability can be understood in terms of the amplification of dynamically unstable phonon modes.  For a cylindrically-symmetric condensate, the system develops a periodic structure of density shells for $\edd>0$ or disks for $\edd<0$, which become amplified and subsequently collapse.

We applied the TF equations of motion, previously used to model oscillatory and expansion dynamics, to study non-adiabatic collapse of the condensate. We showed that this method is valid providing that zero-point kinetic effects remain small throughout.  This can be ensured by employing a TF condensate initially, and suddenly switching the interactions deep into the collapse regime.  The predictions are in excellent agreement with the full Gross-Pitaevskii equation for the majority of the collapse dynamics.  The TF predictions fail when the condensate develops local density structures and thereby deviates from the TF parabolic density profile.

Our results are consistent with the experiment by Lahaye {\it et
al.} \cite{lahaye}.  There the increase of $\edd$ was sufficiently
slow to ensure that global collapse dominates the system.  However,
by changing the interactions more suddenly, it should be possible to
induce local collapse of the condensate.  A rich variety of
structures can be formed depending upon the orientation of the
dipoles and the aspect ratio of the trap. These structures are
related to those that have been predicted to occur in the ground
state, but during local collapse they become amplified. Such
transient structures could be observed experimentally by Bragg
scattering of light.

\appendix

\section{Gaussian variational solutions}
When the TF approximation is not valid (e.g. for weak interactions),
it can be appropriate to employ a gaussian ansatz.  We consider a
cylindrically-symmetric gaussian ansatz of the form \cite{yi,koch},
\begin{eqnarray}
\psi_{\rm g}&=&\sqrt{\frac{\kappa N}{\pi^{3/2} l_x^3}} \exp
\left[-\frac{1}{2l_{x}^2}\left(x^2+y^2+\kappa^2 z^2\right)
\right],\label{eqn:ga}
\end{eqnarray}
where $l_x$ is the transverse width and $\kappa$ is the aspect
ratio. The energy of this ansatz is,
\begin{eqnarray}
\frac{E}{N\hbar
\omega_x}=\frac{1}{2l_x^2}+\frac{\kappa^2}{4l_x^2}+\frac{l_x^2}{2}+\frac{\gamma^2
l_x^2}{2\kappa^2}+\frac{\kappa k_{s}\left[1-\varepsilon_{\rm
dd}f(\kappa)\right]}{\sqrt{2\pi}l_x^3}.
\end{eqnarray}
Variational solutions are then obtained by minimising this with respect to $l_x$ and $\kappa$.

\section{Numerical solution of the dipolar GPE}
A split-step
fast fourier transform method is employed to evolve $\psi({\bf r})$
on a three-dimensional spatial grid \cite{ronen}, typically with $64^3$ points. The dipolar atomic potential in {\it k}-space is
given by,
\begin{equation}
 \tilde{U}_{\rm dd}(k)=\frac{4\pi C_{\rm dd}}{3}\left(\frac{3k_z^2}{k^2}-1\right),
\end{equation}
where $k_x$, $k_y$ and $k_z$ are the cartesian wavevectors and
$k^2=k_x^2+k_y^2+k_z^2$. By combining Eq.~(\ref{eqn:Udd}) with the
convolution theorem, the dipolar mean-field potential is given by
$\Phi_{\rm dd}({\bf r})=\mathcal{F}^{-1}[ {\tilde U}_{dd}(k) {\tilde
n}(k)]$, where the inverse Fourier transform, denoted by
$\mathcal{F}^{-1}$, is performed numerically using a standard fast
Fourier transform algorithm. For pancake and spherical systems we
employ the corrections to ${\tilde U}_{dd}(k)$ outlined in
Ref.~\cite{ronen}. Ground-state solutions of the dipolar GPE are
found by propagation in imaginary time $t \rightarrow -it$.  From a
suitable initial guess and with renormalisation at each time step,
the GPE converges to the ground state of the system.

To monitor the size of the BEC we evaluate $\langle z^2\rangle$ and
$\langle x^2\rangle$ and relate them to the TF radii $R_z$ and $R_x$
via,
\begin{eqnarray}
\langle z^2\rangle= \int d{\bf r}z^2 |\psi|^2=\frac{N}{7}R^2_{z}
\nonumber\\
\langle x^2\rangle=\int d{\bf r} x^2|\psi|^2=\frac{N}{7}R^2_{x}
 \label{size}
\end{eqnarray}
To compare the numerical widths directly with the TF scaling parameters $\lambda_z(t)$ and $\lambda_x(t)$ we define,
\begin{eqnarray}
\lambda_z(t)=\sqrt{\frac{\langle z^2(t)\rangle}{ \langle z^2(0)\rangle}}
\\
\lambda_x(t)=\sqrt{\frac{\langle x^2(t)\rangle}{\langle x^2(0)\rangle}}
\end{eqnarray}
Furthermore, the total energy $E_{\rm tot}$ of the dipolar BEC is
numerically evaluated from the GPE energy functional \cite{dalfovo}
and given by,
\begin{equation}
 E_{\rm tot}=\int_V \left(\frac{\hbar^2}{2m}|\nabla \psi|^2+V_{\rm ext}|\psi|^2+\Phi_{\rm dd}|\psi|^2+\frac{g}{2}|\psi|^4 \right).\label{eq:Efun}
\end{equation}
The terms in the integrand correspond to, from left to right, the kinetic energy $E_{\rm k}$, potential energy $E_{\rm p}$, dipolar interaction energy $E_{\rm d}$ and {\it s}-wave interaction energy $E_{\rm s}$.  Note that the zero-point kinetic energy is the kinetic energy associated with variations in the real part of $\psi$.  Expressing $\psi=\psi_{\rm r}+i\psi_{\rm i}$, where $\psi_{\rm r}$ and $\psi_{\rm i}$ are real quantities, the $|\nabla\psi|^2$-term in Eq.~(\ref{eq:Efun}) can be decomposed as $|\nabla\psi|^2=[\nabla \psi_{\rm r}]^2+[\nabla \psi_{\rm i}]^2$.  The zero-point kinetic energy can thus be defined as,
\begin{equation}
 E_{\rm zp}=\int_V \frac{\hbar^2}{2m}[\nabla \psi_{\rm r}]^2.
\end{equation}


\begin{thebibliography}{99}
\bibitem{varepsilon} The definition $\varepsilon_{\rm dd}$ arises naturally from an analysis of the
frequencies of collective excitations in a homogeneous dipolar BEC:
for $\varepsilon_{\mathrm{dd}}
>1$ instabilities can occur in the hydrodynamic
limit \cite{goral}.
\end{thebibliography}
\end{document}